\documentclass[prl,10pt,floatfix,amsmath,twocolumn,showpacs,superscriptaddress]{revtex4-1}
\usepackage[ansinew]{inputenc}

\usepackage{verbatim}
\usepackage{epsfig}
\usepackage{pst-all}
\usepackage{color}
\usepackage{dcolumn}
\usepackage{amsmath}
\usepackage{amsthm}
\usepackage{bm}
\usepackage{layout}
\usepackage{float}
\usepackage{txfonts}
\usepackage{amsfonts}
\usepackage{amssymb}%
\usepackage[ansinew]{inputenc}
\usepackage{amsmath}
\usepackage{amsthm}
\usepackage{float}
\usepackage{amsfonts}
\usepackage{amssymb}

\newtheorem{lemma}{Lemma}
\newtheorem{thm}[lemma]{Theorem}

\begin{document}

\title{Detecting genuine multipartite correlations in terms of the rank of coefficient matrix}

\author{Bo Li}
\email{libobeijing2008@gmail.com}
\affiliation{Department of Mathematics and Computer, Shangrao Normal University, Shangrao 334001, China}
\affiliation{Institute of Physics, Chinese Academy of Sciences, Beijing 100190, China}

\author{Leong Chuan Kwek}
\affiliation{Center for Quantum Technologies, National University of Singapore, 2 Science Drive 3, Singapore 117543\\
and National Institute of Education and Institute of Advanced Studies and Institute of Advanced Studies, Nanyang Technological University, Singapore, 637616.}

\author{Heng Fan}
\email{hfan@iphy.ac.cn}
\affiliation{Institute of Physics, Chinese Academy of Sciences, Beijing 100190, China}

\date{\today}

\begin{abstract}
We propose a method to detect genuine quantum correlation for arbitrary quantum state in terms
of the rank of  coefficient matrices associated with the pure state. We then derive a
necessary and sufficient condition for  a quantum state to possess  genuine correlation, namely that all corresponding coefficient matrices
have rank larger than one. We demonstrate an approach to decompose the genuine quantum
correlated state with high rank coefficient matrix into the form of product states with no genuine quantum
correlation for pure state.
\end{abstract}

\pacs{03.67.Mn, 03.65.Ud}

\maketitle

\section{Introduction}
Quantification of multipartite entanglement has remained an outstanding but important
challenge in quantum information science. Besides applications in quantum information science,
multipartite states have served as crucial resources for a myriad of quantum information processing
tasks ranging from quantum cryptography and secret sharing \cite{hillery,tittel} to quantum simulation \cite{lloyd},
measurement-base quantum computation \cite{briegel} and high precision metrology \cite{giovannetti}.

While bipartite entanglement is well understood, genuine multipartite entangled states are less studied.
Genuine $n$-partite entanglement needs entanglement of all $n$ constituent systems. It should not be separable for any partition
of the system. Moreover, it turns out that genuine multipartite entanglement are needed in several quantum information processes.
Thus it is important to detect not just multipartite entanglement \cite{papp,krammer,Toth,miyake,seevinck,Vicente} but also genuine multipartite entanglement.
However, detection of genuine multipartite entangled state experimentally however has proved elusive.

Bennett\cite{Bennett} \emph{et al.} have recently proposed some reasonable
postulates for measures of genuine multipartite correlations.
They gave the following postulates or definitions:

\emph{Definition 1.} A state of $n$ particles has genuine $n$-partite correlations
if it is not a product state in every bipartite cut.

\emph{Definition 2.} A state of $n$ particles has genuine $k$-partite correlations if there exists a
$k$-particle subset whose reduced state has genuine $k$-partite correlations.

\emph{Definition 3.} A state has degree of correlations equal to $n$ if there exists a subset
of $n$ particles which has genuine $n$-partite correlations and there does not exist a subsets
of $m$ particles which has genuine $m$-partite correlations for any $m>n$.

The above definition is slightly different from the genuine multipartite entanglement, i.e.,
for pure state, a state has genuine $n$-partite correlation if and only if it has genuine entanglement.
For mixed state, however, a state has genuine $n$-partite entanglement if it is not a mixture of pure
states that do not have genuine  $n$-partite entanglement. Comparing the definitions, it is not
difficult to find that the set of  genuine correlated states that contains the set of genuine entanglement states
for mixed states.

Very recently, Giorgi\cite{Giorgi} \emph{et al.} introduced a  measure of genuine total, classical, and
quantum correlations based on the use of the relative
entropy to quantify the distance between the studied and the corresponding benchmark states; Rulli\cite{rulli,xu,Okrasa} \emph{et al.}
introduce some measure of global quantum discord and geometry discord.
Beside these,  very little is known to us on multipartite quantum correlation.
In this paper, we provide a necessary and sufficient
condition for detecting genuine multipartite correlations for arbitrary $n$-qubit quantum system
 in terms of the coefficient matrices of the pure state.
Then we demonstrate an approach to decompose a state  into the product of genuine
correlated states.

\section{Coefficient matrix and genuine multipartite correlation for pure $n$-qubit  state.}
 We first review the construction of coefficient matrix
as introduced in Ref.\cite{LDF11b,LDF12}.
 Let $|\psi \rangle_{1\cdots n}=\sum_{i=0}^{2^{n}-1}a_{i}|i\rangle $
be an $n$-qubit pure state. We rewrite $|\psi \rangle_{1\cdots n}$ as
 $|\psi \rangle_{1\cdots n}  = \sum_{j=0}^{2^{n-\ell}-1}\sum_{i=0}^{2^{\ell}-1}a_{ij}|i \rangle\otimes|j \rangle$.
 A coefficient matrix associated to the state $|\psi\rangle_{1\cdots n}$ is given by
 \begin{eqnarray}
 C_{1\cdots \ell,(\ell+1)\cdots n}(|\psi \rangle_{1\cdots n})=(a_{ij})_{2^{\ell}\times2^{n-\ell}}. \label{ma-1}
\end{eqnarray}
We remark that the splitting of $\ell:(n-\ell)$ is arbitrary so $\ell$ can take any possible values.
 For more general case, we designate a permutation  $\{q_{1},q_{2},\cdots ,q_{n}\}$ of
$\{1,2,\cdots ,n\}$,  and  $C_{q_{1}\cdots q_{\ell},q_{\ell+1}\cdots
q_{n}}(|\psi \rangle_{1\cdots n})$ be the  coefficient matrix corresponding to the permutation constructed from  $C_{12\cdots
\ell,\ell+1\cdots n}(|\psi \rangle_{1\cdots n})$ in Eq. (\ref{ma-1}).
For convenience,  we omit the subscripts
$q_{\ell+1},\cdots, q_n$ and simply write $C_{q_{1}\cdots q_{\ell}}$,
whenever the column bits are clear from the context.
 If $|\psi \rangle_{1\cdots n}$ is a bipartite product state, i.e.,
$|\psi \rangle_{1\cdots n}=|\phi \rangle_{q_{1}\cdots q_{\ell}}\otimes|\phi' \rangle_{q_{\ell+1}\cdots
q_{n}}$, by the construction of the coefficient matrix,
we have the rank of $C_{q_{1}\cdots q_{\ell}}(|\psi \rangle_{1\cdots n})$ must be $r(C_{q_{1}\cdots q_{\ell}}(|\psi \rangle_{1\cdots n}))=1$.  Together with definition 1 , we obtain a very easy criterion
to detect the genuine multipartite correlation for pure $n$-qubit state.
\begin{thm}
  A pure  $n$-qubit  state $|\psi \rangle_{1\cdots n} $ has genuine $n$-partite correlations
if and only if  for every permutation  $\{q_{1},q_{2},\cdots ,q_{n}\}$ and $0<\ell<n$,
the rank of the coefficient matrix has $r(C_{q_{1}\cdots q_{\ell}}(|\psi \rangle_{1\cdots n}))\neq1$.
\end{thm}
\par

To show the power of Theorem 1, we first give an example to show that the $n$-qubit symmetric Dicke states $|\ell,n\rangle$ with
$\ell(0<\ell< n)$ excitations \cite{stockton}
\begin{equation}
|\ell,n\rangle =\left( _{\ell}^{n}\right) ^{-1/2}\sum\limits_k
P_{k}|1_{1},1_{2},\cdots, 1_{\ell},0_{\ell+1},\cdots, 0_{n}\rangle,
\label{Dicke}
\end{equation}
have genuine correlations, where $\{P_k\}$ is the set of all distinct permutation of the spins.
It is shown in \cite{LDF11b} that the rank of $|\ell,n\rangle$ is $\ell+1$, for $0<\ell< n$,
we have $r(C_{q_{1}\cdots q_{\ell}}(|\ell,n\rangle))\geq 2$ for any permutation.
By Theorem 1, we have that  any $n$-qubit symmetric Dicke states $|\ell,n\rangle$ with
$\ell(0<\ell<n)$ have genuine correlations, for $\ell=0$ or $\ell=n$, we have $|0,n\rangle$ and $|n,n\rangle$
are both product state.

We now further consider arbitrary Permutation symmetric  pure $n$-qubit state, by using Vieta's formulas, any symmetric  pure $n$-qubit state
can be written  into the combination of symmetric Dicke states\cite{Bastin},
\begin{eqnarray}
 |\psi \rangle_{1\cdots n}=\sum_{\ell=0}^n a_{\ell}|\ell, n\rangle, \label{sysmetry}
\end{eqnarray}
we first state that if there exist $\ell(0<\ell<n)$ such that $a_{\ell}^2\neq a_{\ell+1}a_{\ell-1}$,
then  $|\psi \rangle_{1\cdots n}$ in (\ref{sysmetry}) must have genuine correlation.
In fact, for any permutation,  one selects $i_2\cdots  i_{n-1}$ so that
$\ell-1$ bits are equal to $1$ and the rest of bits  are equal to $0$.
The coefficients of  $|0i_2\cdots  i_{n-1}1\rangle$ and
$|1i_2\cdots  i_{n-1}0\rangle$ are $a_{\ell}$,
  the coefficient of  $|0i_2\cdots  i_{n-1}0\rangle$ is $a_{\ell-1}$,
and that of $|1i_2\cdots  i_{n-1}1\rangle$ is $a_{\ell+1}$. We can  find that
the foregoing coefficients constitute a $2\times 2$ nonzero minor of the coefficient matrix. Therefore, we have   $r(C_{q_{1}\cdots q_{\ell}}(|\psi \rangle_{1\cdots n}))\geq 2$, and the state has genuine multipartite correlation.
If we have $a_{\ell}^2= a_{\ell+1}a_{\ell-1}$ for all $\ell(0<\ell<n)$, that is $\frac{a_0}{a_1}=\frac{a_1}{a_2}=\cdots=\frac{a_{n-1}}{a_n}\equiv \alpha$,
then $|\psi \rangle_{1\cdots n}$ is given as
\begin{eqnarray}
 |\psi \rangle_{1\cdots n}& = & a_n\sum_{\ell=0}^n  \alpha^{n-\ell} |\ell, n\rangle\nonumber\\
  &=& a_n(\alpha|0\rangle +|1\rangle )\otimes\cdots\otimes(\alpha|0\rangle +|1\rangle ),
 \label{sysmetryequ}
\end{eqnarray}
which is a product state. In summary, we find that an arbitrary symmetric pure $n$-qubit states have genuine correlations
except the   state is in  one of the following three cases: $|0, n\rangle$,  $|n, n\rangle$, $|\psi \rangle_{1\cdots n}$ in Eq. (\ref{sysmetryequ}).
Thus, we have completely describe the genuine correlations for arbitrary symmetric pure $n$-qubit states.

Using Theorem 1, we can decompose a pure state into a
product of genuine correlated states. For any pure state $|\psi \rangle_{1\cdots n}$ ,
if there exists a permutation $\{q_{1},q_{2},\cdots ,q_{n}\}$ and $0<\ell<n$ such that $r(C_{q_{1}\cdots q_{\ell}})=1$,
then we have the decomposition  $|\psi \rangle_{1\cdots n}=|\phi \rangle_{q_{1}\cdots q_{\ell}}\otimes|\phi' \rangle_{q_{\ell+1}\cdots
q_{n}}$. We can further decompose $|\phi \rangle_{q_{1}\cdots q_{\ell}},|\phi' \rangle_{q_{\ell+1}\cdots
q_{n}}$ into a product of some other states, and eventually obtain that
$|\psi \rangle_{1\cdots n}$ be the product of some genuine correlated states.
It is not difficult to discern that for any pure state, regardless of the permutation, the decomposition to genuine
correlated states is unique\cite{explain}. By definitions 2 and 3, for any pure state, we immediately have
the following theorem.
\begin{thm}
 Suppose that $|\psi \rangle_{1\cdots n}$ has the decomposition
$|\psi \rangle_{1\cdots n}=|\psi \rangle^{(m_1)}\otimes\cdots\otimes\psi \rangle^{(m_M)}$,
and $|\psi \rangle_{1\cdots n}$ has genuine $m_1, \cdots, m_M$-partite correlations, then the
degree of correlations is $\max\{m_1, \cdots, m_M\}$,
here $|\psi \rangle^{(m_1)}, \cdots, \psi \rangle^{(m_M)}$ are states of $m_1, \cdots, m_M$-particles
 genuine correlated states, and $n=m_1+\cdots+m_M$.
\end{thm}
\par

\section{Genuine multipartite correlation for arbitrary  $n$-qubit mixed state}
We next consider the detection of genuine correlations for mixed state. Suppose a $n$-particle state $\rho$ has a bipartite cut
 $\rho=\rho_1\otimes\rho_2$, with $\rho_1=\sum_{i=1}^r\lambda_i|\psi_i \rangle\langle\psi_i|$,
 $\rho_2=\sum_{j=1}^s\mu_j|\phi_j \rangle\langle\phi_j|$ be the spectral decomposition. The density matrices
$\rho_1,\rho_2$ are the $n_1,n_2$-particle density matrix respectively($n_1+n_2=n$),
and $\lambda_i\mu_j, |\psi_i \rangle\otimes|\phi_j \rangle$ are the eigenvalues and
eigenvectors of $\rho$, the rank of $\rho$ is $r\times s$. We can regard $\rho_1$ as a reduced state of a pure
$(n_1+r)$-qubit state
\begin{eqnarray}
 |\psi \rangle=\sum_{i=1}^r\sqrt{\lambda_i}|\psi_i \rangle\otimes|\underbrace{0\cdots1_i\cdots0}_{r}\rangle. \label{psip}
\end{eqnarray}
Similarly, $\rho_2$ can be seen as a reduced state of a pure
$(n_2+s)$-qubit state
\begin{eqnarray}
 |\phi \rangle=\sum_{j=1}^s\sqrt{\mu_j}|\phi_j \rangle\otimes|\underbrace{0\cdots1_j\cdots0}_{s}\rangle. \label{phip}
\end{eqnarray}
Thus, $\rho$ can be seen as a reduced state of a pure $(n_1+n_2+r+s)$-qubit state
\begin{eqnarray}
 |\Phi \rangle  & = &\sum_{i=1}^r\sqrt{\lambda_i}|\psi_i \rangle|\underbrace{0\cdots1_i\cdots0}_{r}\rangle\otimes\nonumber\\
 & & \sum_{j=1}^s\sqrt{\mu_j}|\phi_j \rangle |\underbrace{0\cdots1_j\cdots0}_{s}\rangle\nonumber\\
  & = &|\psi \rangle\otimes|\phi \rangle. \label{Phip}
\end{eqnarray}

For a generic mixed state $\rho$, we provide the following process to
detect the genuine correlations. We first give the spectral decomposition of $\rho$,
\begin{eqnarray}
 \rho=\sum_{i=1}^R\lambda_i|\Phi_i \rangle\langle\Phi_i|, \label{rhop}
\end{eqnarray}
for every factors  $a, b$  of $R$ supply  $R=a\times b$,
 we construct a pure
$(n+a+b)$-qubit state
\begin{eqnarray}
 |\Phi \rangle^{ab} & = &\sum_{i=1}^a\sum_{j=1}^b\sqrt{\lambda_{(i-1)\times b+j}}|\Phi_{(i-1)\times b+j} \rangle\otimes
 |\underbrace{0\cdots1_i\cdots0}_{a}\rangle\nonumber\\
 & & \otimes|\underbrace{0\cdots1_j\cdots0}_{b}\rangle.\label{Phirp}
\end{eqnarray}

One sees that $\rho$ is a reduced state of $|\Phi \rangle^{ab}$ by tracing out
the last $(a+b)$-partite.
If $\rho$ has a bipartite cut with rank $a\times b$, Eq. (\ref{Phirp}) must take the form of
Eq. (\ref{Phip}), and since $ |\Phi \rangle$ in Eq. (\ref{Phip}) is a product state,
we have a permutation $\{q_{1},q_{2},\cdots ,q_{n}\}$ of
$\{1,2,\cdots ,n\}$, such that the rank of coefficient matrix of $ |\Phi \rangle^{ab}$ has
$r(C_{q_{1}\cdots q_{\ell} n+1\cdots n+a}(|\Phi \rangle^{ab}))=1$,  here $0<\ell<n$. On the contrary, if there exists a permutation
 $\{q_{1},q_{2},\cdots ,q_{n}\}$ and $a,b$
such that the rank of the coefficient matrix of $ |\Phi \rangle^{ab}$ in Eq. (\ref{Phirp}) gives $r(C_{q_{1}\cdots q_{\ell} n+1 \cdots n+a}(|\Phi \rangle^{ab}))=1$, then $|\Phi \rangle^{ab}$
can be rewritten as  a $(q_{1}\cdots q_{\ell} n+1 \cdots n+a)\times(q_{\ell+1}\cdots q_{n} n+a+1 \cdots  n+a+b)$-bipartite
product state, by tracing out the $(a+b)$-partite,  we obtain that $\rho$ is a product of  $(q_{1}\cdots q_{\ell})$-partite
by $(q_{\ell+1}\cdots q_{n})$-partite. Thus,
together with definition 1 given by Bennett \emph{et al.}, we have proved the following theorem.
\begin{thm}
  Suppose a  $n$-particle state $\rho$ has rank $R$,  the spectral decomposition of  $\rho$ is given by
$\rho=\sum_{i=1}^R\lambda_i|\Phi_i \rangle\langle\Phi_i|$,
for every factors $a,b$  of $R$ supply $R=a\times b$, construct  $(n+a+b)$-particle pure state
 $|\Phi \rangle^{ab}$  in the form of  Eq. (\ref{Phirp}), then $\rho$ has
 genuine $n$-particle correlation if and only if for every $a,b$,
  permutation $\{q_{1},q_{2},\cdots ,q_{n}\}$, and $0<\ell<n$, the rank of  the coefficient matrix of $ |\Phi \rangle^{ab}$ gives rise to
 $r(C_{q_{1}\cdots q_{\ell}n+1\cdots n+a}(|\Phi \rangle^{ab}))\neq1$.
\end{thm}
\par
 Theorem 3 provides us with a natural way to detect genuine $k$-partite correlations for a state $\rho$ of $n$ particles, and also its degree of correlations.
 According to the definition 2, 3, we need only check if there exists a subset
 $\{q_{1}\cdots q_{k}\}$  of $\{1,2,\cdots ,n\}$ such that each reduced
 density matrix with $\{q_{1}\cdots q_{k}\}$ particles, $\rho_{q_{1}\cdots q_{k}}$, has genuine multipartite correlations to confirm that $\rho$ has genuine $k$-particle correlations, and the maximum $k$ value is just the degree of correlations.

 We show a  flow chart in Fig. 1 to demonstrate the decomposition of an arbitrary $n$-qubit quantum state into a product of  genuine $k$ correlated states.  From the flowchart, we see that we can use an algorithmic approach to decompose a quantum state into a product of genuine $k$ correlated states, and obtain
 its degree of correlations.

We now provide some illustrative examples concerning the usefulness of Theorem 3. Consider the $n$-particle  state
 $\rho=p|\ell, n \rangle\langle\ell,n|+(1-p)|\ell',n \rangle\langle\ell',n|$,  the rank of $\rho$ is $r(\rho)=2$, and
 $|\Phi \rangle^{ab}$  in   Eq. (\ref{Phirp}) given by
\begin{eqnarray}
 |\Phi \rangle^{ab}  = \sqrt{p}|\ell,n \rangle|1\rangle|10\rangle+\sqrt{1-p}|\ell',n \rangle|1\rangle|01\rangle.\label{Phirpsysmma}
\end{eqnarray}
Since $\ell\neq\ell'$, for any $\{q_1,\cdots,q_k\}$, we can select the term $|\alpha_1\cdots\alpha_n \rangle$ in
$|\ell, n \rangle$ and $|\beta_1\cdots\beta_n \rangle$ in $|\ell', n \rangle$
which are different on $\{q_1,\cdots,q_k\}$-particles, notice that the coefficients of
$|\alpha_1\cdots\alpha_k\beta_{k+1}\cdots\beta_n \rangle|1\rangle|10\rangle$,
$|\beta_{1}\cdots\beta_k\alpha_{k+1}\cdots\alpha_n \rangle|1\rangle|01\rangle$ are zero,
we have $C( |\Phi \rangle^{ab}_{q_1\cdots q_kn+1})$ has a nonzero $2\times2$ minor.
Therefore $r(C( |\Phi \rangle^{ab}_{q_1\cdots q_kn+1}))\geq2$ for any $0<k<n$,
which means $\rho$ has genuine multipartite correlation.
As a corollary, we can also show that
 $\rho=p|GHZ \rangle\langle GHZ|+(1-p)|W \rangle\langle W|$ has genuine multipartite correlation.

\begin{figure}\centerline{
\includegraphics[width=5cm]{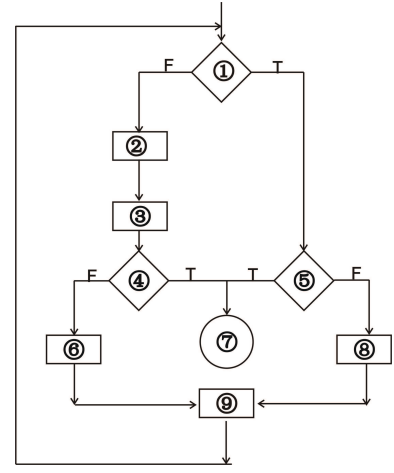}}
\caption{ (1) Verify the state is pure or not; (2) Decompose $\rho$ into spectral decomposition;
(3) For every factors $a,b$ of the rank $R$ such that $a\times b=R$, construct $(n+a+b)$-partite
pure state $ |\Phi \rangle^{ab}$ in Eq. (\ref{Phirp}); (4) For every permutation $\{q_{1},q_{2},\cdots ,q_{n}\}$
and $0<\ell<n$, verify the rank of the coefficient matrix $r(C_{q_{1}\cdots q_{\ell} n+1\cdots n+a}(|\Phi \rangle^{ab}))=1$ or not;
(5) For every permutation $\{q_{1},q_{2},\cdots ,q_{n}\}$
and $0<\ell<n$, verify the rank of the coefficient matrix $r(C_{q_{1}\cdots q_{\ell}}(|\Phi \rangle))=1$ or not;
(6) Decompose $\rho$ into the product of $(q_{1}\cdots q_{\ell})\times(q_{\ell+1}\cdots q_n)$ bipartite state
$\rho=\rho_1\otimes\rho_2$; (7) $\rho$ or  $ |\Phi \rangle$ must be genuine correlated state;
(8) Decompose $|\Phi \rangle$ into the product of $(q_{1}\cdots q_{\ell})\times(q_{\ell+1}\cdots q_n)$ bipartite state
$|\Phi \rangle=|\phi \rangle_{q_{1}\cdots q_{\ell}}\otimes|\phi' \rangle_{q_{\ell+1}\cdots
q_{n}}$; (9) Input $\rho_1,\rho_2,|\phi \rangle_{q_{1}\cdots q_{\ell}},|\phi' \rangle_{q_{\ell+1}\cdots
q_{n}}$ into (1) respectively. ``T'' represent true, and ``F'' is false.
}\label{lct}
\end{figure}

\medskip

\section{Relationship between the higher rank (larger than one) of the coefficient matrix and no genuine correlated state.}
It is shown by Li \emph{et al.} \cite{LDF12}  that the rank of
any coefficient matrix is invariant under stochastic local operations and
classical communication (SLOCC).
For the state with coefficient matrix of higher rank, we give the following theorem which indicates that
the  rank is a kind of measure of genuine multipartite correlations.
\begin{thm}
  For a $n$ particle pure state $|\psi \rangle_{1\cdots n}$ , and a  permutation $\{q_{1},q_{2},\cdots ,q_{n}\}$,
if the rank of the coefficient matrix $r(C_{q_{1}\cdots q_{\ell}}(|\psi \rangle_{1\cdots n}))=k$,($k>1$),
then $|\psi \rangle_{1\cdots n}$  can be expressed as a  sum of  $k$ $(q_{1}\cdots q_{\ell})\times(q_{\ell+1}\cdots q_{n})$
bipartite product states, each of them has no genuine correlation.
\end{thm}
\par
\begin{proof}
  We first consider the pure state. Suppose the  permutation
 $\{q_{1},q_{2},\cdots ,q_{n}\}$ and $\ell (0<\ell<n)$ such that
 $r(C_{q_{1}\cdots q_{\ell}}(|\psi \rangle_{1\cdots n}))=k$,  we rewrite $|\psi \rangle_{1\cdots n}=\sum_{i=0}^{2^{\ell}-1}\sum_{j=0}^{2^{n-\ell}-1}a_{ij}|i\rangle
 \otimes|j\rangle$, where $|i\rangle$ represents the $q_{1}\cdots q_{\ell}$ qubit and $|j\rangle$ represents the $q_{\ell+1}\cdots q_{n}$ qubit.
 The coefficient matrix $C_{q_{1}\cdots q_{\ell}}(|\psi \rangle_{1\cdots n})$
 is given as $(a_{ij})_{2^{\ell}\times2^{n-\ell}}$.
Denote each column vector of matrix  as $\beta_j$
\begin{eqnarray*}
 \beta_j=(a_{0j},\cdots,a_{2^{\ell}-1j})^T,j=0,\cdots,2^{n-\ell}-1,
\end{eqnarray*}
where $T$ is the transpose of the vector, for convenience, we suppose $\beta_0,\cdots,\beta_{k-1}$
is linear independent, and the rest of columns are the linear combination of the preceding $k$ columns,
\begin{eqnarray*}
 \beta_j=\sum_{v=0}^{k-1}t_{vj}\beta_v, j=k,\cdots,2^{n-\ell}-1,
\end{eqnarray*}
which means
\begin{eqnarray}
 a_{ij}=\sum_{v=0}^{k-1}t_{vj}a_{iv},\label{aij}
\end{eqnarray}
where $i=0,\cdots,2^{\ell}-1, j=k,\cdots,2^{n-\ell}-1$.

We   decompose  $|\psi \rangle_{1\cdots n}$ into two terms,
$|\psi \rangle_{1\cdots n}=|\psi \rangle_1+|\psi \rangle_2$, where
$|\psi \rangle_1  = \sum_{j=0}^{k-1}(\sum_{i=0}^{2^{\ell}-1}a_{ij}|i \rangle)\otimes|j \rangle$, and
$|\psi \rangle_2  = \sum_{j=k}^{2^{n-\ell}-1}(\sum_{i=0}^{2^{\ell}-1}a_{ij}|i \rangle)\otimes|j \rangle$.
Inserting (\ref{aij}) into $|\psi \rangle_2 $, we have
\begin{eqnarray}
 |\psi \rangle_2=\sum_{v=0}^{k-1}\sum_{i=0}^{2^{\ell}-1}\sum_{j=k}^{2^{n-\ell}-1}t_{vj}a_{iv}
 |i \rangle\otimes|j \rangle,\label{Phirpp}
\end{eqnarray}
change the letter $v$ as $j$, and $j$ as $v$, we have that
\begin{eqnarray}
 |\psi \rangle_2 &= &\sum_{j=0}^{k-1}\sum_{i=0}^{2^{\ell}-1}\sum_{v=k}^{2^{n-\ell}-1}t_{jv}a_{ij}
 |i \rangle\otimes|v \rangle\nonumber\\
 &= &\sum_{j=0}^{k-1}(\sum_{i=0}^{2^{\ell}-1}a_{ij}
 |i \rangle)\otimes(\sum_{v=k}^{2^{n-\ell}-1}t_{jv}|v \rangle),\label{Phirprp}
\end{eqnarray}
therefore,
\begin{eqnarray}
 |\psi \rangle_{1\cdots n}=\sum_{j=0}^{k-1}(\sum_{i=0}^{2^{\ell}-1}a_{ij}
 |i \rangle)\otimes(|j \rangle+\sum_{v=k}^{2^{n-\ell}-1}t_{jv}|v \rangle).\label{Phirprpr}
\end{eqnarray}
Eq.(\ref{Phirprpr}) means that $ |\psi \rangle$ is a sum of $k$ bipartite product states, which has no
genuine correlation. It is shown that the sum of $k$ rank one matrix must has
rank no more than $k$, we obtain that
$|\psi \rangle_{1\cdots n}$  can not be expressed as a  sum of  less than $k$ $(q_{1}\cdots q_{\ell})\times(q_{\ell+1}\cdots q_{n})$
bipartite product states.
\end{proof}

\section{How the algorithm works}
To show how our algorithm works, we consider further
some examples. We next study an example of the entanglement swapping. Four Bell states are given as
\begin{eqnarray}
 |\beta_{xy} \rangle=\frac{|0,y\rangle+(-1)^x|1,\overline{y}\rangle}{\sqrt{2}},\label{bells}
\end{eqnarray}
where $x,y=0,1$ and $\overline{y}$ is the negation of $y$. We now construct a four parties entanglement
state $|\Phi\rangle_{AB|CD}$ as follows
\begin{eqnarray}
 |\Phi\rangle_{AB|CD}=\frac{1}{2}\Sigma_{x,y=0}^1|\beta_{xy} \rangle_{AB}\otimes|\beta_{xy} \rangle_{CD}.\label{phiabcd}
\end{eqnarray}
We can find that the coefficient matrix $C(|\Phi\rangle_{AC})$ and $C(|\Phi\rangle_{AD})$ are given by
\begin{eqnarray}
C(|\Phi\rangle_{AC})=C(|\Phi\rangle_{AD})=\frac{1}{2}\left(
                                                        \begin{array}{cccc}
                                                          1 & 0 & 0 & 1 \\
                                                          0 & 0 & 0 & 0 \\
                                                          0 & 0 & 0 & 0 \\
                                                          1 & 0 & 0 & 1 \\
                                                        \end{array}
                                                      \right)
,\label{bells}
\end{eqnarray}
which is rank $1$.  Therefore, by Theorem 1, $|\Phi\rangle_{AB|CD}$ has no genuine multipartite  correlation and can be expressed as
a product of $AC|BD$ and $AD|BC$ bipartite cut. In fact,
\begin{eqnarray*}
 |\Phi\rangle_{AB|CD}=|\beta_{00} \rangle_{AC}\otimes|\beta_{00} \rangle_{BD}=|\beta_{00} \rangle_{AD}\otimes|\beta_{00} \rangle_{BC},
\end{eqnarray*}
is the condition for the so called entanglement swapping similar as teleportation, that is often invoked in quantum information processes so that a long-distance entangled state is generated
from the short-distance entangled states.

We next use our method to prove that Smolin state \cite{smolin} has genuine multipartite correlation. The Smolin state is given by
\begin{eqnarray}
 \rho=\frac{1}{4}\Sigma_{x,y=0}^1|\beta_{xy} \rangle_{AB}\langle\beta_{xy}| \otimes|\beta_{xy} \rangle_{CD}\langle\beta_{xy}| ,\label{rhoabcd}
\end{eqnarray}
Eq. (\ref{rhoabcd}) is actually the spectral decomposition of $\rho$, and the  rank of $\rho$ is $4$, which
can be  a factorization of $2\times2$ and $1\times4$, thus, the purification of $\rho$ to $(4+2+2)-$qubit pure state is given as
\begin{eqnarray}
 |\Phi\rangle^{22}=\frac{1}{2}\Sigma_{x,y=0}^1|\beta_{xy} \rangle_{AB} \otimes|\beta_{xy} \rangle_{CD}\otimes|x,\overline{x}\rangle_{EF}\otimes|y,\overline{y} \rangle_{GH}.\label{phyab}
\end{eqnarray}
Explicitly, we have,
\begin{eqnarray}
 & &|\Phi\rangle^{22}=\nonumber\\
 &  &\frac{1}{4}((|00 \rangle+|11 \rangle)_{AB}\otimes(|00 \rangle+|11 \rangle)_{CD}\otimes|01 \rangle_{EF}\otimes|01 \rangle_{GH}\nonumber\\
 & &+(|01 \rangle+|10 \rangle)_{AB}\otimes(|01 \rangle+|10 \rangle)_{CD}\otimes|01 \rangle_{EF}\otimes|10 \rangle_{GH}\nonumber\\
 &  &+(|00 \rangle-|11 \rangle)_{AB}\otimes(|00 \rangle-|11 \rangle)_{CD}\otimes|10 \rangle_{EF}\otimes|01 \rangle_{GH}\nonumber\\
 &  &+(|01 \rangle-|10 \rangle)_{AB}\otimes(|01 \rangle-|10 \rangle)_{CD}\otimes|10 \rangle_{EF}\otimes|10 \rangle_{GH}).\nonumber\\
 \label{phyabse}
\end{eqnarray}
It is an $8$-qubit pure state. By Theorem 3, we first check whether the rank of the following $7$ matrices
$$C(|\Phi\rangle_{AEF}^{22}),C(|\Phi\rangle_{BEF}^{22}),C(|\Phi\rangle_{CEF}^{22}),C(|\Phi\rangle_{DEF}^{22}),$$
$$C(|\Phi\rangle_{ABEF}^{22}),C(|\Phi\rangle_{ACEF}^{22}),C(|\Phi\rangle_{ADEF}^{22})$$ are
equal to one or not (it is not difficult to find that $r(C(|\Phi\rangle_{AEF}^{22}))=r(C(|\Phi\rangle_{AGH}^{22}))=r(C(|\Phi\rangle_{BCDEF}^{22}))$).
It will be tedious for direct verifying since they are $8\times32$ or $16\times16$ order matrices.
In the following, we only need to show that there is a nonzero $2\times2$ minor for each matrix.
We first give a detailed construction of the nonzero $2\times2$ minor of $C(|\Phi\rangle_{AEF}^{22})$.
From (\ref{phyabse}), we select 2 terms and write them into $AEF|BCDGH$ bipartite cut,
$$|00\rangle_{AB}|00\rangle_{CD}|01\rangle_{EF}|01\rangle_{GH}=|001\rangle_{AEF}|00001\rangle_{BCDGH},$$
$$|11\rangle_{AB}|00\rangle_{CD}|01\rangle_{EF}|01\rangle_{GH}=|101\rangle_{AEF}|10001\rangle_{BCDGH}.$$
By (\ref{phyabse}) the coefficients of $|001\rangle_{AEF}|00001\rangle_{BCDGH}$ and $|101\rangle_{AEF}|10001\rangle_{BCDGH}$
are $\frac{1}{4}$, we further construct two terms, $|101\rangle_{AEF}|00001\rangle_{BCDGH}$ and $|001\rangle_{AEF}|10001\rangle_{BCDGH}$,
which  lie on the cross of the row and column of the forging two terms and  do not belong to the terms of (\ref{phyabse}),
 the coefficients are $0$. Therefore, the four coefficients constitute a nonzero
$2\times2$ minor of $C(|\Phi\rangle_{AEF}^{22})$, which means $r(C(|\Phi\rangle_{AEF}^{22}))\geq2$.
It is shown that for any permutation and combination of different parties, we can always find a nonzero minor of the coefficient,
the submatrices we select are all $\frac{1}{4}\begin{pmatrix}
                                               1 & 0 \\
                                               0 & 1 \\
                                             \end{pmatrix}
$.

The purification of $\rho$ to $(4+1+4)-$qubit pure state is given as
\begin{eqnarray}
 |\Phi\rangle^{14} & = &
 \frac{1}{4}((|00 \rangle+|11 \rangle)_{AB}\otimes(|00 \rangle+|11 \rangle)_{CD}\otimes|1 \rangle_{E}\otimes|1000 \rangle_{FGHI}\nonumber\\
 & &+(|01 \rangle+|10 \rangle)_{AB}\otimes(|01 \rangle+|10 \rangle)_{CD}\otimes|1 \rangle_{E}\otimes|0100 \rangle_{FGHI}\nonumber\\
 &  &+(|00 \rangle-|11 \rangle)_{AB}\otimes(|00 \rangle-|11 \rangle)_{CD}\otimes|1 \rangle_{E}\otimes|0010 \rangle_{FGHI}\nonumber\\
 &  &+(|01 \rangle-|10 \rangle)_{AB}\otimes(|01 \rangle-|10 \rangle)_{CD}\otimes|1 \rangle_{E}\otimes|0001 \rangle_{FGHI}),\nonumber\\
 \label{phyabseD}
\end{eqnarray}
for any permutation and combination of particles $``ABCD"$. Since Smolin state
is highly symmetric, we can always find a nonzero $2\times2$ minor
of the coefficient matrix with combining $``E"$,
 by Theorem 3, we have that the Smolin state has genuine multipartite correlation.

\section{Summary and discussion.}
In this paper we provide an efficient  method for detecting genuine multipartite correlations
of arbitrary $n$-qubit system in terms of the rank of  coefficient matrices.
The necessary and sufficient condition for a state with genuine correlation
can be related to the rank of the coefficient matrix of a pure state, which is shown to be
invariant under SLOCC. The proposed measure
satisfies those general postulates raised by Bennett \emph{et al.} in Ref.\cite{Bennett}.

We would like to remark that the proposed method is essentially algorithmic.
Thus one can follow the algorithm systematically to check for the
genuine multipartite correlation. We have also provided several interesting
examples for pure or mixed state cases to illustrate the power of the approach, namely  symmetric
pure states, a state related to entanglement swapping, a mixed state of W state
and GHZ state (which has genuine multipartite correlation), and the Smolin state (which also has genuine multipartite correlation).

\acknowledgements

We thank Dafa Li for helpful discussion. This work is partially supported by ``973'' program (2010CB922904), NSF of China and the
National Research Foundation and Ministry of Education, Singapore.


\begin{thebibliography}{99}

\bibitem{hillery} M. Hillery, V. Buzek and A. Berthiaume, Phys. Rev. A {\bf 59},
1829 (1999).
\bibitem{tittel} W. Tittel, H. Zbinden and N. Gisin, Phys. Rev. A, {\bf 63} 042301 (2001).
\bibitem{lloyd} S. Lloyd, Science {\bf 273}, 1073 (1996).
\bibitem{briegel} R. Raussendorf, H. J. Briegel, Phys. Rev. Lett. 86, 5188 (2001).
\bibitem{giovannetti} V. Giovannetti, S. Lloyd and L. Maccone, Phys. Rev. Lett. 96, 010401 (2006).

\bibitem{papp} S.B. Papp, K. S. Choi, H. Deng, P. Lougovski, S. J. van Enk, H. J. Kimble1,
  Science. {\bf 324}, 764 (2009).
\bibitem{krammer}  P. Krammer, H. Kampermann, D. Bru{\ss}, R. A. Bertlmann, L. C. Kwek, and C. Macchiavello,
 Phys. Rev. Lett. {\bf 103}, 100502 (2009).

\bibitem{Toth} G. T\'{o}th and O. G\"{u}hne, Phys. Rev. Lett. \textbf{94}, 060501 (2005).

\bibitem{miyake} A. Miyake and H. J. Briegel, Phys. Rev. Lett. {\bf 95}, 220501 (2005).
\bibitem{seevinck} M. Seevinck and J. Uffink, Phys. Rev. A \textbf{78},  032101 (2008).
\bibitem{Vicente} J. I. de Vicente, and M. Huber , Phys. Rev. A \textbf{84},  062306 (2011).
\bibitem{Bennett} C. H. Bennett, D. Grudka, M. Horodecki, P. Horodecki, and R. Horodecki,
Phys. Rev. A \textbf{83}, 012312 (2011).
\bibitem{Giorgi} G. L. Giorgi, B. Bellomo, F. Galve, and R. Zambrini, Phys. Rev. Lett. \textbf{\bf 107}, 190501 (2011).
\bibitem{rulli}
C. C. Rulli and M. S. Sarandy , Phys. Rev. A \textbf{84}, 042109 (2011).
\bibitem{xu}
J. Xu, J. Phys. A: Math. Theor. 45 (2012) 405304.
\bibitem{Okrasa} M. Okrasa, Z. Walczak, Europhys. Lett. \textbf{96}, 60003 (2011)
\bibitem{LDF11b} X. Li and D. Li, Phys. Rev. Lett. \textbf{\bf 108}, 180502 (2012).
\bibitem{LDF12} X. Li and D. Li, arXiv:1201.2229.
\bibitem{Bastin}T. Bastin, S. Krins, P. Mathonet, M. Godefroid, L. Lamata, and E. Solano, Phys. Rev. Lett.\textbf{\bf 103}, 070503 (2009).
\bibitem{stockton} J. K. Stockton, J. M. Geremia, A. C. Doherty, and H. Mabuchi,
Phys. Rev. A \textbf{67}, 022112 (2003).
\bibitem{smolin}J. A. Smolin, Phys. Rev. A {\bf 63}, 032306 (2001).
\bibitem{explain} Suppose  $|\psi \rangle_{1\cdots n}=|\phi \rangle_{q_{1}\cdots q_{\ell}}\otimes|\phi' \rangle_{q_{\ell+1}\cdots
q_{n}}=|\varphi \rangle_{p_{1}\cdots p_{\ell'}}\otimes|\varphi' \rangle_{p_{\ell'+1}\cdots p_{n}}$, then
we use the form $|\varphi \rangle_{p_{1}\cdots p_{\ell'}}\otimes|\varphi' \rangle_{p_{\ell'+1}\cdots p_{n}}$ trace out the
$q_{1}\cdots q_{\ell}$-particles is also a pure state,  which means that $|\psi \rangle_{1\cdots n}$ can be further
decompose to a product of the intersection of $q_{1}\cdots q_{\ell}$ and $p_{1}\cdots p_{\ell'}$-particles, here
$q_{1}\cdots q_{n}$, $p_{1}\cdots p_{n}$  are permutations of $1,\cdots,n$.

\end{thebibliography}
\end{document}